# A Broken Translational Symmetry State in an Infinite-Layer Nickelate

Matteo Rossi[1], Motoki Osada[1,2], Jaewon Choi[3], Stefano Agrestini[3], Daniel Jost[1], Yonghun Lee[1,4], Haiyu Lu[1,5], Bai Yang Wang[1,5], Kyuho Lee[1,5], Abhishek Nag[3], Yi-De Chuang[6], Cheng-Tai Kuo[7], Sang-Jun Lee[7], Brian Moritz[1], Thomas P. Devereaux[1,2,8], Zhi-Xun Shen[1,4,5,8], Jun-Sik Lee[7], Ke-Jin Zhou[3], Harold Y. Hwang[1,4,8] and Wei-Sheng Lee[1*]

[1]Stanford Institute for Materials and Energy Sciences, SLAC National Accelerator Laboratory, Menlo Park, California 94025, USA

[2]Department of Materials Science and Engineering, Stanford University, Stanford, California 94305, USA

[3]Diamond Light Source, Harwell Campus, Didcot OX11 0DE, United Kingdom

[4]Department of Applied Physics, Stanford University, Stanford, California 94305, USA

[5]Department of Physics, Stanford University, Stanford, California 94305, USA

[6]Advanced Light Source, Lawrence Berkeley National Laboratory, Berkeley, California 94720, USA

[7]Stanford Synchrotron Radiation Lightsource, SLAC National Accelerator Laboratory, Menlo Park, California 94025, USA

[8]Geballe Laboratory for Advanced Materials, Stanford University, Stanford, California 94305, USA

* Corresponding author: leews@stanford.edu

**A defining signature of strongly correlated electronic systems is the existence of competing phases with similar ground state energies, resulting in a rich phase diagram [1]. While in the recently discovered nickelate superconductors [2-9], a high antiferromagnetic exchange energy has been reported, which implies the existence of strong electronic correlations [10], signatures of competing phases have not yet been observed. Here, we uncover a charge order (CO) in infinite-layer nickelates $La_{1-x}Sr_xNiO_2$ using resonant x-ray scattering across the Ni $L$-edge. In the parent compound, the CO arranges along the Ni-O bond direction with an incommensurate wave vector (0.344±0.002, 0) r.l.u., distinct from the stripe order in other nickelates [11-13] which propagates along a direction 45º to the Ni-O bond. The CO resonance profile indicates that CO originates from the Ni $3d$ states and induces a parasitic charge modulation of La electrons. Upon doping, the CO diminishes and the ordering wave vector shifts toward a commensurate value of 1/3 r.l.u., indicating that the CO likely arises from strong correlation effects and not from Fermi surface nesting.**

The infinite-layer nickelates have emerged as a new manifestation of unconventional superconductivity (SC). Despite the formal similarity with high-temperature cuprate superconductors in crystal structure and valence electron counting [14], the low-energy electronic structure exhibits notable differences. Experiments [15,16] suggest that the infinite-layer nickelates possess a relatively large charge transfer energy, placing it away from a dominant charge transfer regime in the Zaanen-Sawatzky-Allen (ZSA) classification scheme [17]. In addition, x-ray spectroscopies [15], transport measurements [3,4,6,7,9], and density functional theories [18-20] suggest multi-orbital band characteristics involving Fermi surfaces associated not only with the Ni 3$d$ states, but also with the rare-earth 5$d$ states. Interestingly, recent resonant inelastic x-ray scattering (RIXS) measurements revealed the existence of a sizable antiferromagnetic (AFM) exchange interaction between Ni spins via the direct observation of dispersive magnetic excitations that resemble the spin waves in a spin-1/2 AFM square lattice [10]. The high energy scale of the magnetic bandwidth (~200 meV) is consistent with a Mott system in the strong coupling regime (i.e., large onsite Coulomb interaction U). This implies that strong electronic correlations, which are a key ingredient in the cuprate phenomenology [21,22], might be also at play in the nickelate superconductors. It is of great importance to search for another key signature of strong correlations, such as the emergence of competing orders [1], that not only sheds new light on the underlying electronic structure, but also critically establishes the infinite-layer nickelates as a new type of strongly correlated electron system.

To investigate this issue, we employ resonant x-ray scattering near the Ni $L_3$ absorption edge to search for signatures of broken translational symmetry states in the La-based infinite-layer nickelates. Figure 1a displays the RIXS intensity map of undoped LaNiO$_2$ along the ($h$, 0)-direction (i.e., Ni-O bond direction). A clear enhancement of the quasi-elastic signal at an in-plane momentum transfer $\mathbf{Q_{CO}}$ = (±0.344±0.002, 0) r.l.u. is seen (Fig. 1c), while a similar feature is absent along the ($h$, $h$)-direction (i.e., the Ni-Ni bond direction) as shown in Fig. 1b and 1d. The position of $\mathbf{Q_{CO}}$ is distinct from the allowed Bragg reflections of the underlying crystallographic symmetry, suggesting that the scattering of $\mathbf{Q_{CO}}$ is due to the presence of a charge order that breaks the crystallographic translational symmetry. Indeed, as expected in a CO system, a softening of the RIXS-derived phonon dispersion is also observed (Fig. 1a and Extended Data Fig. 1). Note that the CO in LaNiO$_2$ has a relatively short correlation length (~ 40 Å) and propagates along the ($h$, 0)-direction, distinct from the stripe order in other nickel oxides, such as single-layer nickelates La$_2$NiO$_{4+y}$ [11] and La$_{2-x}$Sr$_x$NiO$_4$ [12] and tri-layer La$_4$Ni$_3$O$_8$ [13], which propagate along a direction 45º to the Ni-O bond (*i.e.* along the ($h$, $h$)-direction in the tetragonal unit cell notation). In addition, no anomaly in the magnetic excitations near **Q$_{CO}$** can be resolved (Extended Data Fig. 2), unlike those in striped nickelates [23]. Thus, the short-range CO in LaNiO$_2$ appears to bear more resemblance to the CO in Bi-, Y-, and Hg-based cuprates [24-26].

To gain further insights, we examine the incident energy dependence of the CO peak across representative absorption edges of La and Ni ions using RIXS and resonant soft x-ray scattering (RSXS) (Methods and Extended Data Fig. 3). The CO peak with the same wave vector can be detected when the incident photon energy is tuned to the La $M_5$ (~834.1 eV, Extended Data Fig. 4), La $M_4$ (~850.58 eV) and Ni $L_3$-edges (~852.58 eV) (Fig. 1e). The CO peak intensity reaches a maximum at approximately 0.3 eV below the main Ni $L_3$ absorption peak and rapidly vanishes away from the absorption edges (Fig. 1e and 1g). These observations indicate that both La and Ni states exhibit charge density modulations with an identical periodicity. Interestingly, as shown in Fig. 1f and 1g, by introducing just 4% holes (x = 0.04), the CO almost vanishes at the La $M_4$-edge, whereas the CO resonance remains robust near the Ni $L_3$-edge. This suggests that CO originates from the Ni $3d$ states. The charge modulation at the La sites is induced by that of Ni states via coupling between the Ni and La ions, which appears to weaken rapidly with increasing doping.

The doping evolution of the CO is further investigated using Ni $L_3$-edge RIXS, which provides higher sensitivity to the CO state in the quasi-elastic scattering. Figure 2a,b,c show raw RIXS intensity maps of La$_{1-x}$Sr$_x$NiO$_2$ for x = 0.06, 0.10 and 0.15, respectively, which were used to extract CO peak profiles (Fig. 2d) from the quasielastic region of the RIXS spectra. As shown in Fig. 2d, the CO peak weakens and broadens monotonically with increasing doping and becomes undetectable at x = 0.15. Upon hole doping, the amplitude of the CO order parameter, which is related to the peak area shown in Fig. 2e, reduces, indicative of a smaller charge density variation in the doped compounds. In addition, the broadening of the peak with doping (Fig. 2f) indicates the decrease of correlation length ($\varepsilon = \sigma^{-1}$, where $\sigma$ is the standard deviation obtained from the Gaussian fit) from ~10 unit cells in the undoped compound to ~6 unit cells for x = 0.10. Interestingly, the CO wave vector decreases slightly, moving towards a commensurate value of 1/3 r.l.u. as shown in Fig. 2g, indicating that CO is likely arising from strong correlation effects and not from Fermi surface nesting.

The temperature dependence of CO in $La_{1-x}Sr_xNiO_2$ is summarized in Fig. 3. RSXS scans of undoped $LaNiO_2$ (Fig. 3a and Extended Data Fig. 4) reveal that the peak persists to at least 286 K (the highest achievable temperature of our experimental setup), indicating that the onset temperature of the CO phase is notably higher than room temperature. For x = 0.06, the CO peak gradually weakens with temperature until the signal vanishes within the error bars at around ~ 200 K (Fig. 3b and 3d). For x = 0.10, as the CO signal is beyond the detection limit of RSXS, we thus employ RIXS to reveal its temperature dependence. Figure 3c shows that the peak is already very weak and broad at low temperatures and disappears at 148 K. As summarized in Fig. 3d and 3e, the CO peak gradually weakens with increasing temperature, while the onset temperature lowers upon doping. The persistence of the CO correlation to high temperatures without a sharp onset appears to be consistent with the absence of a clear signature of a phase transition in transport measurements [7]. This is distinct from a conventional charge density wave system, but resembles the short-range CO correlation in Y-, Bi- and Hg-based cuprates which also persists to high temperatures without a sharp onset or notable signatures in transport measurements [27-30].

Based on the temperature and doping dependence data of CO, we propose an updated phase diagram of $La_{1-x}Sr_xNiO_2$, displayed in Fig. 4. Interestingly, the CO regime roughly coincides with the territory where weakly insulating behavior is observed [7], suggesting that they may be related. However, its relationship to SC appears to be puzzling. On the one hand, SC emerges at higher doping as CO diminishing, which may suggest that these two phases are competing. On the other hand, the CO is strongest in the undoped parent compound, and here SC also sets in below ~ 1 K [7], implying that these two may be coupled somewhat. The relationship between CO and SC remains an important open question for future investigations, much as it is under active debate in the cuprates and other materials.

It is intriguing that the CO already occurs in the undoped infinite-layer nickelates, whereas CO should not exist in a half-filled Mott-Hubbard system. This indicates that a finite percentage of holes should already reside in the Ni $3d$ bands in the nominally x = 0 compounds. While we cannot completely rule out that some of the holes were associated with possible oxygen defects in the compounds, the systematic doping evolution of the CO suggests that oxygen defects are unlikely a dominant factor. The holes in the Ni $3d$ states are likely introduced via a self-doping effect whereby charge is transferred via a coupling to the Fermi surface pockets associated with the rare earth $5d$ states [18-20], as also inferred from transport measurements [3,4]. As suggested by a recent theoretical calculation for a coupled rare earth $5d$ and Ni $3d$ two-band model, the rare-earth-Ni coupling induces a self-doping effect and gives rise to a modulation in both rare earth and Ni charge densities with similar periodicities [31]. When the rare earth Fermi surface pocket is depleted by hole doping, only the Ni charge density is modulated [31]. These behaviors are consistent with our experimental finding in La$_{1-x}$Sr$_x$NiO$_2$ (Fig. 1). Moreover, the fact that the parasitic CO in La states is significantly reduced by introducing a small hole doping lends further support to a small Fermi surface pocket of the La states.

While in cuprates the CO emerges from the $3d^9L$ configuration (*L* denotes a hole in the O ligand states) with a strong O 2*p* character, the CO in the infinite-layer nickelates possesses a dominant $3d^8$ character strongly associated with the doped holes [16,32]. This indicates that the CO instability is inherent in the strong correlation of the Ni 3*d* orbitals, without significant ligand contributions. In addition, a CO modulation along the (*h*, 0)-direction would disfavor the formation of long-range antiferromagnetic (AFM) order along the (*h*, *h*)-direction. In other words, the competition between the CO and the AFM correlation might be another reason why long-range AFM order has not been observed [33,34] despite a sizable spin exchange interaction [10]. Finally, it remains an important question regarding whether the CO is ubiquitous in other infinite-layer nickelates. We note that the emergence of CO in the infinite-layer nickelates might be very sensitive to the local environment, such as epitaxial strain, chemical pressure induced by different rare-earth elements, or disorder. Nonetheless, the intriguing interplay between CO, AFM, SC, and the underlying multi-orbital electronic structure establishes the infinite-layer nickelate as a new type of strongly correlated electron system.

## Methods

**Materials and sample characterization.** Thin films of the precursor phase La$_{1-x}$Sr$_x$NiO$_3$ with thickness of ~ 9 nm were grown by pulsed laser deposition on a single-crystalline SrTiO$_3$ (001) substrate. A capping layer of SrTiO$_3$ with thickness of 5 unit cells was grown on top of the film to preserve the crystal structure. The infinite-layer phase La$_{1-x}$Sr$_x$NiO$_2$ was obtained by means of a topotactic reduction process using CaH$_2$ powder. Details on the thin film growth and characterization are reported in Ref. [7].

**Ultrahigh resolution Ni $L_3$-edge RIXS measurements.** Ni $L_3$-edge RIXS measurements were performed at beamline I21 of the Diamond Light Source (United Kingdom). The incident energy was tuned 0.1 eV below the maximum of the Ni $L_3$ absorption spectrum, while the incident photon polarization was orthogonal to the scattering plane (σ polarization), unless otherwise indicated. The combined energy resolution was 36 meV for RIXS spectra measured with σ incident photon polarization and 41 meV for RIXS spectra measured with π polarization. The scattering angle was fixed to 154°. RIXS data are plotted as a function of $\mathbf{q}_{//} = (h, k)$, which is the projection of the momentum transfer $\mathbf{q} = \mathbf{k}_i - \mathbf{k}_f$ onto the NiO$_2$ plane, where $\mathbf{k}_i$ and $\mathbf{k}_f$ are the wave vectors of the incident and scattered photons, respectively. The in-plane momentum transfer is denoted in reciprocal lattice units (r.l.u.), i.e. in units of (2π/a, 2π/b) with a = b = 3.91 Å. In our convention, positive $\mathbf{q}_{//}$ corresponds to grazing emission geometry, while negative $\mathbf{q}_{//}$ corresponds to grazing incidence geometry.

**La $M_5$-edge RIXS measurements.** La $M_5$-edge RIXS measurements were performed at qRIXS endstation at beamline 8.0.1 of the Advanced Light Source (USA). The incident photon polarization was parallel to the scattering plane ($\pi$ polarization). The scattering angle was fixed to 150°. The combined energy resolution was ~ 0.6 eV.

**RSXS measurements.** RSXS measurements were carried out at beamline 13-3 of the Stanford Synchrotron Radiation Lightsource (USA). The incident photon polarization was orthogonal to the scattering plane ($\sigma$ polarization). The scattering angle was fixed to 150°. Rocking scans of the CO peak were obtained by rotating the sample around the $b$ axis ($\theta$-scan). For each $\theta$, an image of 256×1024 pixels was collected using the two-dimensional CCD detector. Then, the signal from each row of 256 pixels was integrated to obtain the final two-dimensional image, whose horizontal axis corresponds to $\theta$ and vertical axis corresponds to detector vertical pixel (Extended Data Fig. 3a). The vertically wide CCD covers the signal of interest from the ($h$, 0) plane near the center of the detector as well as signals from the ($h$, ±$k$) planes that are dominated by a fluorescence background. This allows us to simultaneously record the CO peak profile by integrating the signal over a window of 550 pixels near the center of the detector as well as the fluorescence background by integrating the signal over a window of 124 pixels far away from the detector center (Extended Data Fig. 3a,b).

## Acknowledgments

This work is supported by the U.S. Department of Energy (DOE), Office of Science, Basic Energy Sciences, Materials Sciences and Engineering Division, under contract DE-AC02-76SF00515. We acknowledge the Gordon and Betty Moore Foundation's Emergent Phenomena in Quantum Systems Initiative through grant GBMF9072 for synthesis equipment. We acknowledge Diamond Light Source for time on beamline I21-RIXS under Proposal MM25598 and MM27558. RSXS experiments were carried out at the SSRL (beamline 13-3), SLAC National Accelerator Laboratory, supported by the U.S. DOE, Office of Science, Office of Basic Energy Sciences under contract no. DE-AC02-76SF00515. This research also used resources of the Advanced Light Source, a U.S. DOE Office of Science User Facility under contract no. DE-AC02-05CH11231. D.J. gratefully acknowledges support of the Alexander-von-Humboldt foundation via a Feodor-Lynen postdoctoral fellowship.

## Author contributions

M.R. and W.S.L. conceived the research project and designed the experiments. M.R., H.L., D.J., Y.L., J.C., S.A., A.N., K.-J.Z. and W.S.L. performed RIXS experiments at DLS and discussed the results. M.R., D.J., Y.L., H.L., C.T.K., S.J.L., J.S.L. and W.S.L. performed RSXS experiments at SSRL and discussed the results. M.R., H.L., Y.D.C. and W.S.L. performed RIXS experiments at ALS and discussed the results. M.O., B.Y.W., K.L. and H.Y.H. synthesized and characterized the samples. M.R. and W.S.L. analyzed the data. M.R., Z.X.S., B.M., T.P.D., H.Y.H. and W.S.L. interpreted the results. M.R. and W.S.L. wrote the manuscript with input from all authors.

## Competing interests

The authors declare no competing interests.

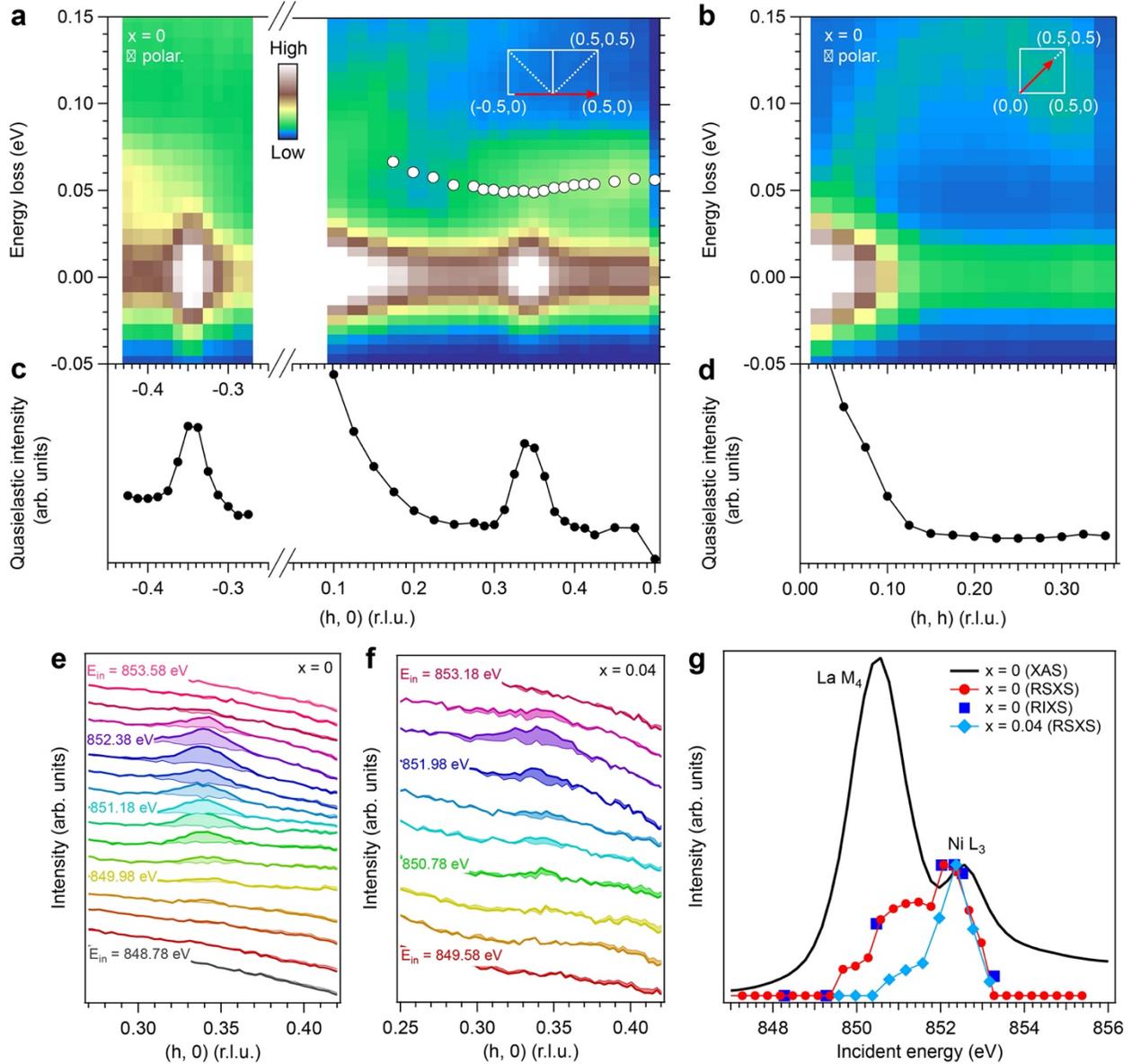

**Figure 1 | Resonant x-ray scattering evidence of translational symmetry breaking in La$_{1-x}$Sr$_x$NiO$_2$. a, b**, Raw Ni $L_3$-edge RIXS intensity map of LaNiO$_2$ measured at 20 K along the (**a**) ($h$, 0)- and (**b**) ($h$, $h$)-direction, as indicated with red arrows in the insets, which depict the first Brillouin zone. White circles represent the phonon dispersion as determined from the fit of the data (see also Extended Data Fig. 1). **c,d**, Intensity of the RIXS quasi-elastic line along the (**c**) ($h$, 0)- and (**d**) ($h$, $h$)-direction, respectively. An enhancement of the quasi-elastic line due to the charge order is evident at **Q**$_{CO}$ = (±0.344, 0) r.l.u. **e,f**, Waterfall plots of RSXS scans of La$_{1-x}$Sr$_x$NiO$_2$ (**e**,

x = 0; **f**, x = 0.04) measured at 26 K along the ($h$, 0)-direction as a function of incident photon energy. The determination of the RSXS signal (thick lines) and background (thin lines) is explained in Methods and Extended Data Fig. 3. **g**, Charge order peak intensity as a function of incident photon energy across the La $M_4$ and Ni $L_3$ edge measured by RSXS and RIXS. X-ray absorption spectrum (XAS) of x = 0 compound is also superimposed. The maximum peak intensity is set to 1 for better comparison.

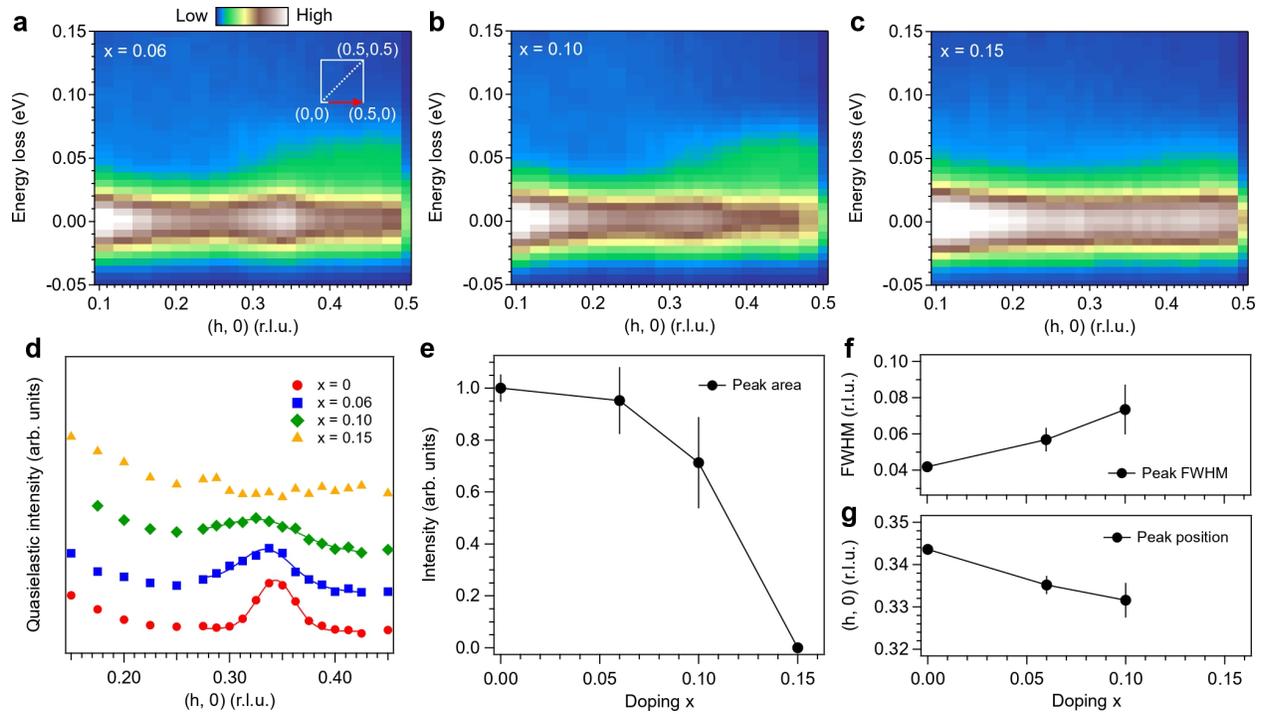

**Figure 2 | Doping dependence of RIXS signal of La$_{1-x}$Sr$_x$NiO$_2$ at 20 K. a, b, c**, Raw RIXS intensity maps of La$_{1-x}$Sr$_x$NiO$_2$ (**a**, x = 0.06; **b**, x = 0.10; **c**, x = 0.15) measured along the (*h*, 0)-direction. The enhancement of the quasi-elastic line at **Q**$_{CO}$ = (0.344, 0) r.l.u. gradually decreases with doping and disappears in the 15%-doped sample. **d**, CO peak profile extracted from the momentum distribution of the quasi-elastic peak intensity in RIXS spectrum. Data are vertically offset for clarity. Solid lines are fit to the peak profile using a Gaussian lineshape and a linear background. **e, f, g**, Doping dependence of the CO peak (**e**, area; **f**, FWHM; **g**, position) determined from the fit of the data in panel **d**.

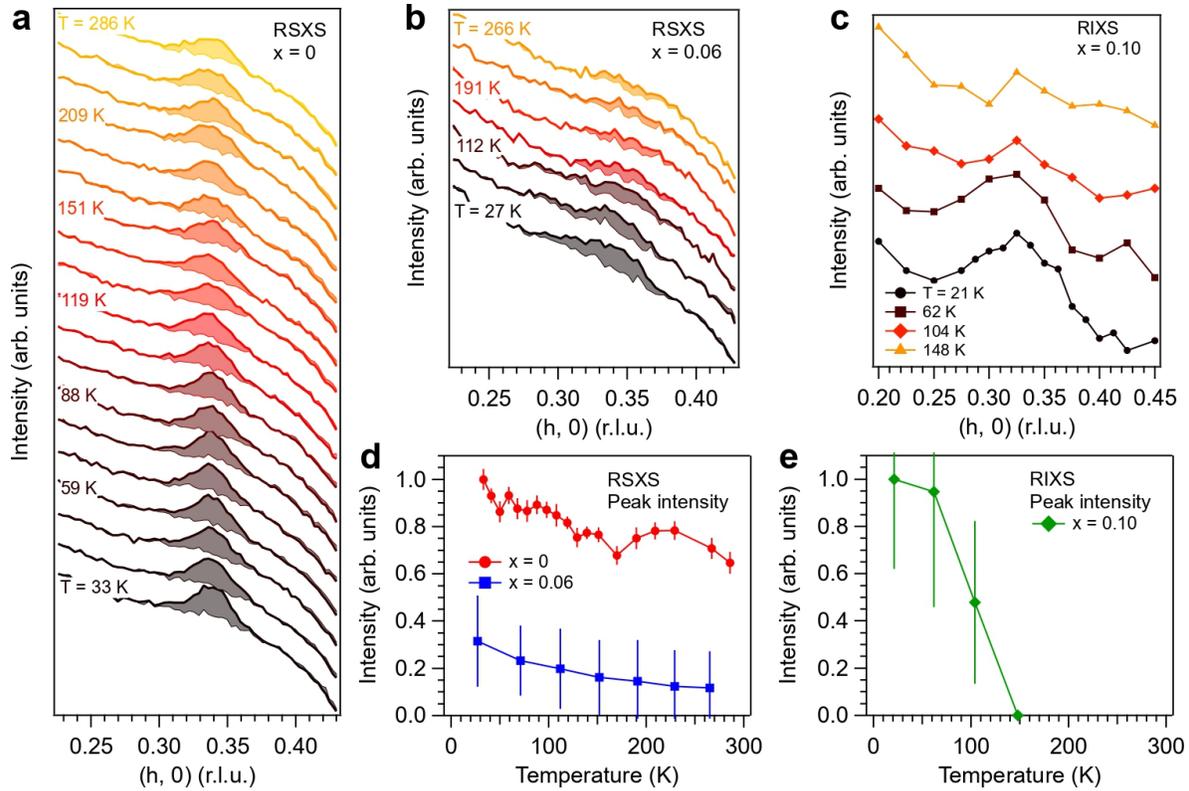

**Figure 3 | Temperature dependence of RSXS and RIXS signal of La$_{1-x}$Sr$_x$NiO$_2$. a**, Waterfall plot showing RSXS scans of LaNiO$_2$ as a function of temperature. The CO peak persists at the highest measured temperature of 286 K. The determination of the RSXS signal (thick lines) and background (thin lines) is explained in Methods and Extended Data Fig. 3. **b**, Temperature dependence of RSXS scans of La$_{1-x}$Sr$_x$NiO$_2$ (x = 0.06). The peak gradually decreases with increasing temperatures. **c**, Temperature dependence of RIXS quasi-elastic signal of La$_{1-x}$Sr$_x$NiO$_2$ (x = 0.10). The CO peak vanishes at the highest measured temperature of 148 K. **d**, Temperature evolution of the CO peak intensity for samples with doping of x = 0 and 0.06. The low-temperature peak intensity of x = 0 has been set to 1. **e**, Temperature dependence of the CO peak intensity for x = 0.10. The peak intensity at low temperature has been normalized to 1.

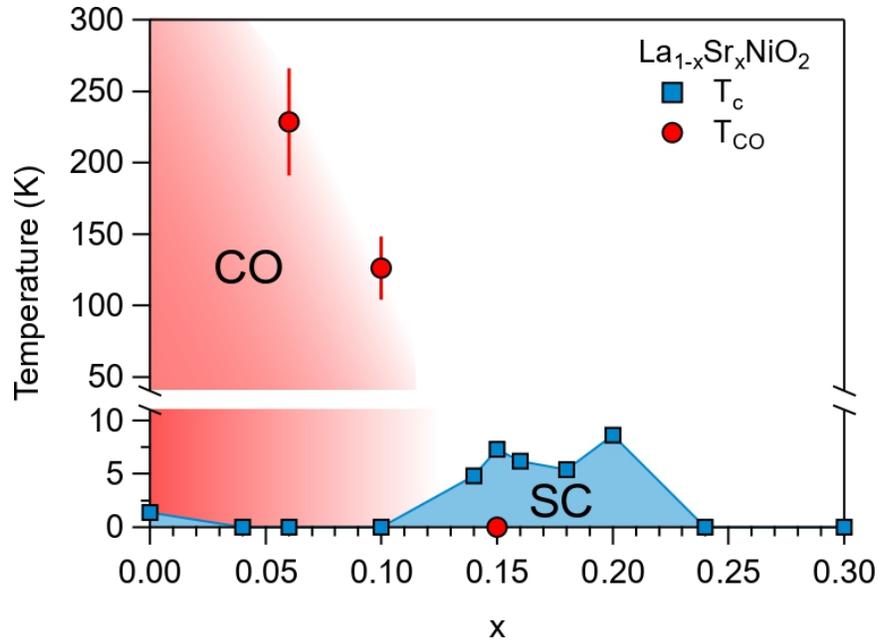

**Figure 4 | Phase diagram of La$_{1-x}$Sr$_x$NiO$_2$.** The red domain represents the region of the phase diagram where charge order (CO) is present and bounded by red circles. Error bars are estimated from the temperature interval used in the measurements. The light blue domain represents the region where the sample is superconducting (SC). Squares display the onset of superconductivity, which is defined as 90% of the resistivity at 20 K (resistivity data are adapted from Ref. [7]).

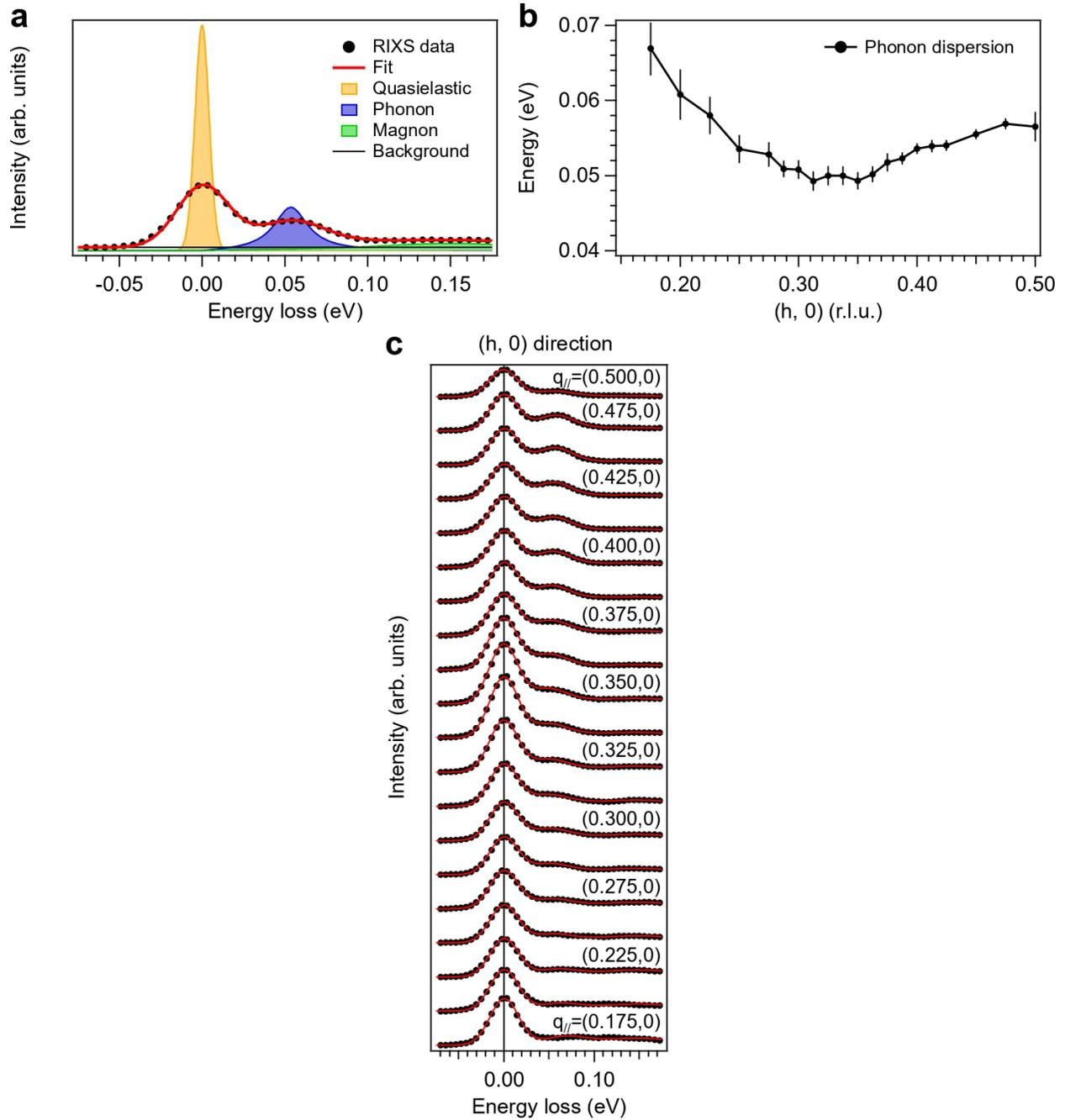

**Extended Data Figure 1 | RIXS phonon softening in LaNiO$_2$ at 20 K and fit of the spectra. a**, Representative RIXS spectrum of LaNiO$_2$ (black circles) at $\mathbf{q}_{//}$ = (0.4, 0) r.l.u. and corresponding fit (red solid line). The fit function includes a Gaussian for the elastic line (orange), anti-symmetrized Lorentzian for phonon (blue) and magnon (green) and a constant background (thin black line). The fit function is convolved with a Gaussian with FWHM = 36 meV. **b**, Phonon

dispersion clearly showing the softening of the RIXS phonon. The phonon dispersion is also shown in Fig. 1a of the main text. **c**, Raw RIXS spectra of LaNiO$_2$ (black circles) with fit superimposed (red lines). The data shown here were taken using σ incident photon polarization to optimize the RIXS phonon signal.

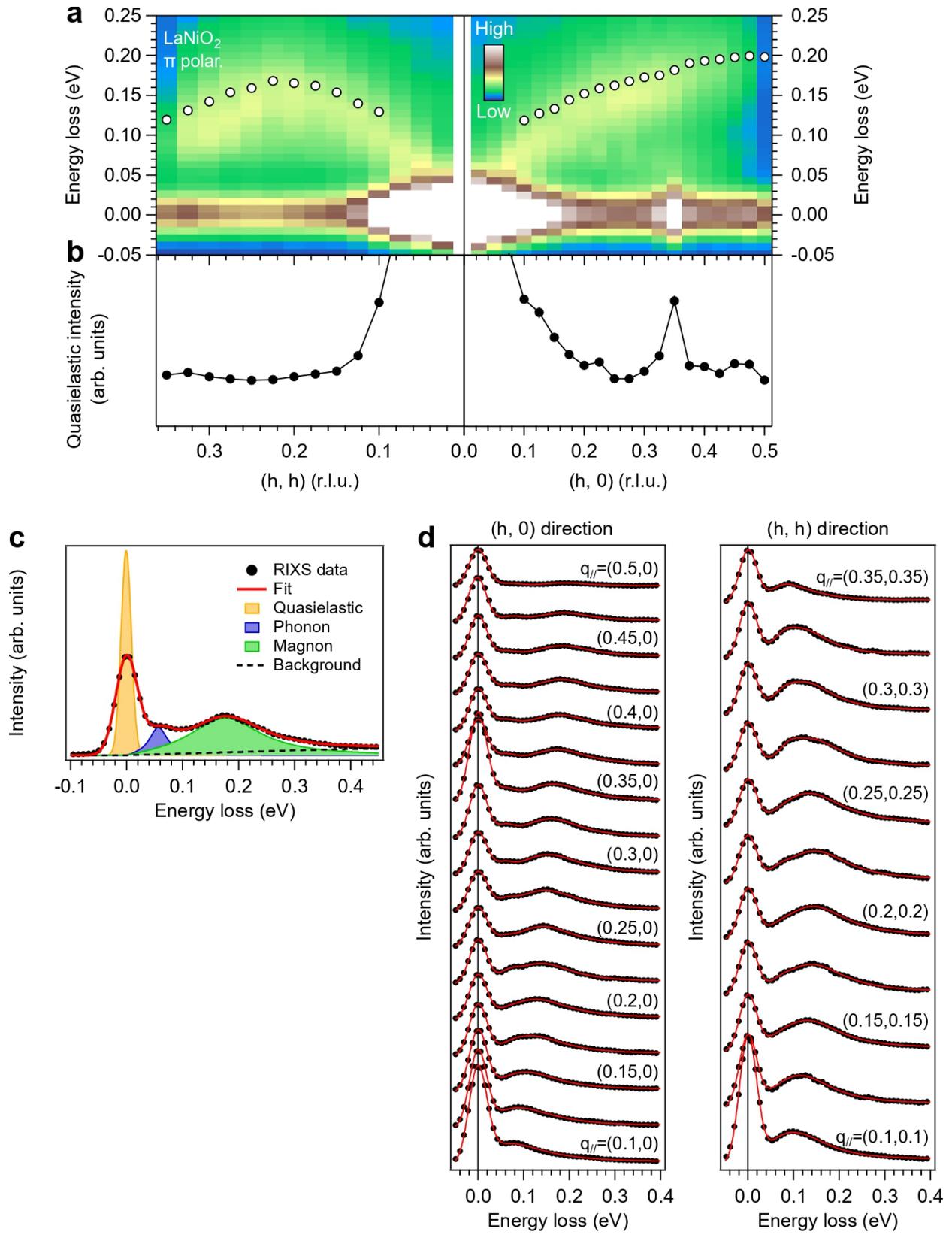

**Extended Data Figure 2 | RIXS intensity map of LaNiO$_2$ at 20 K and fit of the spectra. a**, RIXS intensity map of LaNiO$_2$ measured with π incident photon polarization, which optimize the cross-section of the magnetic excitations. White circles correspond to the undamped mode energies of the magnetic excitation as determined from the damped harmonic oscillator function used for fitting the data. **b**, Intensity of the RIXS quasi-elastic line of LaNiO$_2$ determined from a Gaussian fit. **c**, Representative RIXS spectrum of LaNiO$_2$ (black circles) at **q**$_{//}$ = (0.4, 0) r.l.u. and corresponding fit (red solid line). The fit function includes a Gaussian for the elastic line (orange), anti-symmetrized Lorentzian for phonon (blue) and high-energy background (dashed line) and damped harmonic oscillator function for the magnon (green). The fit function is convolved with a Gaussian with FWHM = 40 meV. **d**, Raw RIXS spectra of LaNiO$_2$ (black circles) measured with π polarization with fit superimposed (red lines).

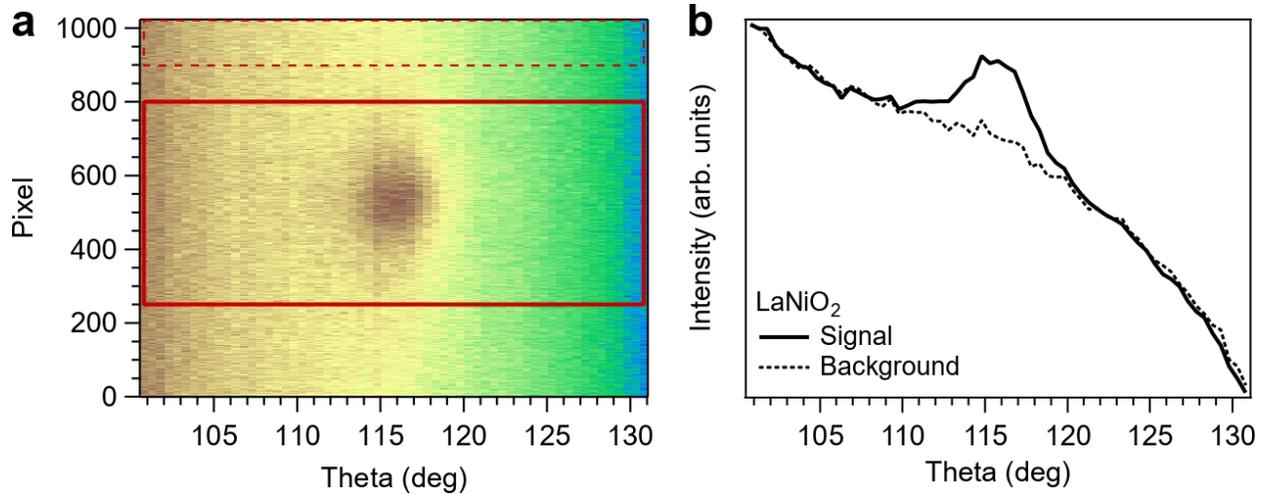

**Extended Data Figure 3 | Extraction of RSXS data from two-dimensional images. a**, Ni $L_3$-edge RSXS scan of LaNiO$_2$ at 33 K obtained by rotating the sample around the *b* axis (θ-scan), which is orthogonal to the scattering plane. Every column is obtained by integrating the rows of the two-dimensional CCD detector collected at a specific θ angle. The CO peak profile is obtained by integrating the columns near the center of the detector (solid red window), while the background is estimated from the fluorescence signal collected at the top of the detector (dashed red window). The region of interest corresponds to k ~ ±0.015 r.l.u., while the top window corresponds to k ~ 0.02-0.03 r.l.u., which is well separated from the CO signal. **b**, The corresponding CO peak profile (thick solid line) and the fluorescence background (thin dashed line). In the main text, the theta angle is converted to in-plane momentum transfer.

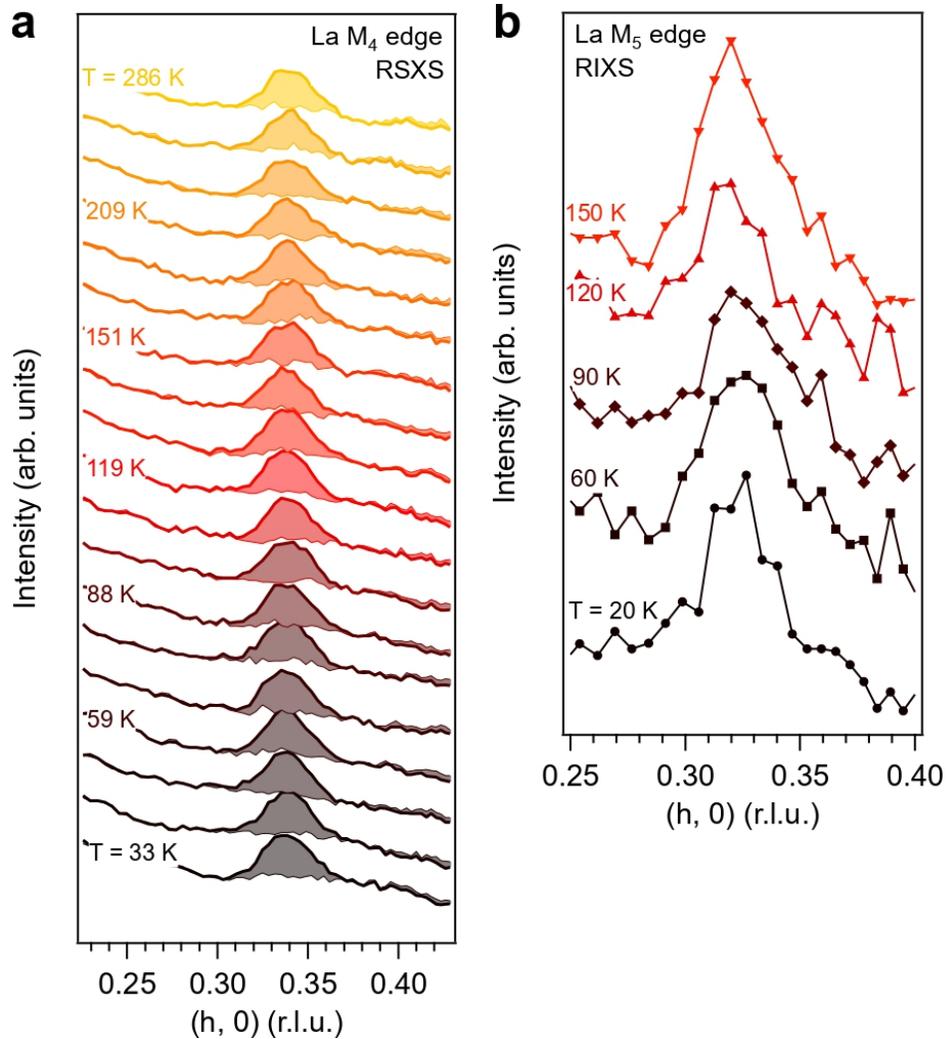

**Extended Data Figure 4 | CO peak profile of LaNiO$_2$ measured at the La $M_4$ and $M_5$ absorption edges. a**, Waterfall plot showing RSXS scans of LaNiO$_2$ as a function of temperature measured with incident photon energy tuned at the La $M_4$ edge (~850.58 eV). **b**, CO peak profile obtained by fitting the RIXS quasielastic line measured at the La $M_5$ edge (~834.1 eV). The CO signal is clearly visible at both absorption edges up to the highest measured temperature.